\documentclass[usenatbib,useAMS]{emulateapj}

\usepackage{natbib}
\usepackage{graphicx,epsfig}
\usepackage{latexsym,amssymb}
\usepackage{bm}
\usepackage{color}

\slugcomment{Submitted to the Astrophysical Journal}

\shorttitle{Redshift distributions from cross-correlations}
\shortauthors{A.E. Schulz}

\begin{document}

\title{Calibrating photometric redshift distributions with cross-correlations}

\author{A.E. Schulz}
\affil{Institute for Advanced Study, Einstein Drive, Princeton NJ
  08540, USA }
\email{schulz@ias.edu}

\date{\today}
\begin{abstract}
The next generation of proposed galaxy surveys will increase the number of galaxies with photometric 
redshift identifications by two orders of magnitude, drastically expanding both the redshift range and 
detection threshold from the current state of the art.  Obtaining spectra for a fair sub-sample of this new data could be cumbersome and expensive.  However, adequate calibration of the true redshift 
distribution of galaxies is vital to tapping the potential of these surveys to illuminate the processes of galaxy evolution, and to constrain the underlying cosmology and growth of structure.  We examine here a promising alternative to direct spectroscopic follow up: calibration of the redshift distribution of photometric galaxies via cross-correlation with an overlapping spectroscopic survey whose members trace the same density field.  We review the theory, develop a pipeline to implement the method, apply it to mock data from N-body simulations, and examine the properties of this redshift distribution estimator.  We demonstrate
that the method is generally effective, but the estimator is weakened by two main factors.  One is that the correlation function of the spectroscopic sample must be measured in many bins along the line of sight, which renders the measurement noisy and interferes with high quality reconstruction of the photometric redshift distribution.  Also, the method is not able to disentangle the photometric 
redshift distribution from evolution in the bias of the photometric sample.  We establish the impact of these factors using our mock catalogs.  We conclude it may still be necessary
to spectroscopically follow up a fair subsample of the photometric survey data.  Nonetheless, it is 
significant that the method has been successfully implemented on mock data, and with further
refinement it may appreciably decrease the number of spectra that will be needed to calibrate future surveys.
\end{abstract}


\section{Introduction}\label{sec:intro}
It is essential to calibrate the true redshift distribution of galaxies in a photometric survey if the survey 
is to be utilized to its full potential.  One application of survey data that requires a detailed understanding
of the distribution of galaxies is weak gravitational lensing tomography.  
The shearing of the shapes of distant galaxies via weak gravitational lensing is a powerful cosmological probe that can be used to study the distribution of dark matter, the nature of dark energy, the formation of large scale structures in the universe, as well as fundamental properties of elementary particles and potential modifications to the general theory of relativity (recent studies include \cite{2008ARNPS..58...99H}, \cite{2009MNRAS.395..197T}, \cite{2009A&A...497..677K}, \cite{2009PhRvD..79b3520I}).   
Cosmic shear measurements statistically examine
minute distortions in the orientations of high redshift galaxies, whose shapes have been sheared by intervening dark matter structures.   Although weak gravitational lensing provides only an integrated 
measure of the intervening density field, using source populations at different redshifts permits some degree of three dimensional reconstruction, known as tomography.  The distortions are small (at the 1\% level) and the intrinsic orientation of the source galaxies is unknown, thus large galaxy samples are 
required to map the density field and probe the growth of density fluctuations with precision.   Existing
cosmic shear measurements have already constrained the amplitude of the dark matter fluctuations 
at the 10\% level \cite{2004MNRAS.347..895H} \cite{2004ApJ...605...29R} \cite{2005MNRAS.359.1277M} \cite{2006A&A...452...51S} and there are many exciting galaxy survey
proposals that will increase the available number of source galaxies by two orders of magnitude 
including DES, DUNE, Euclid, LSST, PanStarrs, SNAP, and Vista.  

Because these large galaxy surveys will have photometric rather than spectroscopic redshift identifications, the community has carefully attended to fine tuning the calibration of the photometric redshifts, minimizing biases and catastrophic errors \cite{2008MNRAS.386..781M},
\cite{2008MNRAS.390..118L}, \cite{2009MNRAS.398.2012F}, \cite{2009AJ....138...95X}, 
\cite{2009arXiv0908.4085G}.  Unlike experiments that use the galaxy positions to directly trace the underlying dark matter distribution, such as baryon acoustic oscillation studies, weak lensing analyses do not require a precise redshift identification for each individual source galaxy.  It is sufficient to accurately determine the redshift distribution of the sources.  
However, lensing measurements are extremely sensitive both to error and bias in the source 
distribution \cite{2002PhRvD..65f3001H}.  Attaining an accurate source distribution will be crucial if weak lensing measurements are to be competitive with other cosmological probes in constraining the 
cosmological parameters. 

Another example where calibration of the true distribution of galaxies may be essential is in using the abundances and clustering of different galaxy populations to connect galaxies at late times to their potential progenitors at early times (as in e.g. \cite{2009ApJ...696..620C}).  Such studies also utilize 
the luminosity functions of galaxies in each redshift slice, and sometimes divide these into different 
rest-frame color bins.  To avoid potential systematics in inferences made about galaxy evolution, 
it will be necessary to know if some 
fraction of the population in a given photometric redshift bin is actually living at a different redshift, especially if there is an asymmetry in such errors that depends on color.  

Recently, an alternative approach to attaining an accurate source redshift distribution has been proposed in 
\cite{2008ApJ...684...88N}.   This method is similar to cross-correlation techniques used in 
\cite{1985MNRAS.212..657P} and
\cite{2006ApJ...644...54M}, and the idea has also been studied theoretically in 
\cite{2006ApJ...651...14S} and \cite{2009arXiv0902.2782B}.   A similar technique was used in 
\cite{2009A&A...493.1197E} to check the redshift distribution for interlopers. Similar in spirit, the
analysis of \cite{2009arXiv0910.2704Q} uses close angular pairs of photometric galaxies to constrain the photometric errors without the use of a spectroscopic sample.

The cross-correlation method determines the photometric redshift distribution by 
utililizing the cross correlation of the galaxies in the photometric sample
with an overlapping spectroscopic sample that traces the same underlying density field.
One advantage of this approach is that the spectroscopic sample used to calibrate the photometric 
redshift distribution can be comprised of bright rare objects such as quasars or 
Luminous Red Galaxies (LRGs)
whose spectra are relatively easy to obtain, and 
indeed may already exist in legacy data.   Spectra could also be obtained for emission line 
galaxies (ELGs), which are easy to follow up but may not represent a fair subsample.  Another 
advantage is that catastrophic redshift errors in the photometry do not systematically bias the 
redshift distribution estimate, they merely contribute to the noise. 

The cross-correlation method
makes use of two observables, the line-of-sight projected angular cross-correlation between the photometric and spectroscopic samples $w_{ps}(\theta)$, and the three dimensional autocorrelation function of the spectroscopic sample $\xi_{ss}(r)$.   By postulating a simple proportionality between the autocorrelation function of the spectroscopic objects and the three dimensional cross-corelation function between the two samples 
$\xi_{ps}(r) \propto \xi_{ss}(r)$, it is potentially possible to infer a very accurate redshift distribution for the photometric sample.  This assumption is guaranteed to be valid if the spectroscopic sample is a sub-sample of the photometric population, but may be problematic if the two sets of tracers have different bias functions with respect to the dark matter. 

In this paper we develop a pipeline to apply the cross-correlation method.  In section \ref{sec:theory} we review the theory and explain how the method works.  We highlight its strengths and examine potential drawbacks and systematic effects.  As a proof of concept, in section \ref{sec:sims} we use the halo model 
to populate N-body simulations with mock photometric and spectroscopic galaxy data to quantify the properties of this redshift distribution estimator.  We examine the extent to which different bias functions interfere with the reconstruction of the true distribution of photometric galaxies.
In section \ref{sec:disc} we discuss inherent tradeoffs, 
outline outstanding theoretical questions, and draw our conclusions.  We leave a more 
detailed discussion of error propagation to the appendix.  

\section{Theory}\label{sec:theory}
The goal of this paper is to demonstrate via numerical simulations that two spatially overlapping
samples, one photometric and one spectroscopic, can be combined to infer the redshift distribution
of the photometric sample to very high accuracy.  The redshift distribution is 
\begin{eqnarray}
\phi_p(z)=\frac{dN_p}{dz\,d\Omega}\,\left[ \int_0^\infty 
   \frac{dN_p}{dz\,d\Omega} \, dz \right]^{-1}
\end{eqnarray}
where $\frac{dN_p}{dz\,d\Omega}$ is the
number of photometric galaxies per unit redshift, per
steradian, and the the quantity in brackets is the the total number of galaxies 
(per steradian) in the sample, ensuring that $\phi_p(z)$ integrates to one.  
If a survey is divided into a number of redshift bins $z_i$, then $\phi_p(z_i)$ 
gives the fraction of the total number of galaxies that live in the $i^{th}$ bin.
Suppose we observe the angular cross-correlation function  
between all the photometric 
galaxies and the spectroscopic galaxies in a particular bin $z_i$.  This  angular cross 
correlation function
is related to the photometric redshift distribution that we are attempting to calibrate.
\begin{eqnarray}\label{eqn:wpsdef}
w_{ps}(\theta,z_i)=\int_0^\infty \xi_{ps}(r(z,z_i,\theta))\,\phi_p(z) \, dz
\end{eqnarray}
Here $\xi_{ps}(r)$ is the three dimensional cross correlation function between the entire photometric
sample and the spectroscopic galaxies that live in bin $i$, which is not observable because 
the redshifts of the photo-z sample are not known to sufficient accuracy to measure it.  The key
assumption of the cross correlation method is that $\xi_{ps}(r)\propto\xi_{ss}(r)$, where 
$\xi_{ss}(r)$, the 3D autocorrelation function of the spectroscopic calibrators, is observable.   
This is a reasonable assumption because on large (linear) scales, 
both $\xi_{ps}(r)$ and $\xi_{ss}(r)$ are related to the underlying dark matter power spectrum 
$\Delta_{\rm lin}^2(k)$ as
\begin{eqnarray}
\xi_{ss}(r) \approx \int_0^\infty \frac{dk}{k} b_s^2\Delta_{\rm lin}^2(k) j_0(kr) \\
\xi_{ps}(r) \approx \int_0^\infty \frac{dk}{k} b_s b_p\Delta_{\rm lin}^2(k) j_0(kr)   
\end{eqnarray}
where $b_s$ and $b_p$ are the linear biases of the spectroscopic and photometric samples
and $j_0$ is a spherical Bessel function.  

In asserting the proportionality, it is implicitly assumed that these biases are scale independent
and that they evolve similarly with redshift.  The former assumption is valid unless the correlation 
functions are being measured on scales smaller than $\sim 1$ Mpc/$h$, while the latter 
assumption may in fact present some difficulty unless the spectroscopic objects are a fair 
subsample of the photometric population.  This is because in real life, the photometric sample
may be apparent magnitude limited.  Thus, the population of galaxies being examined at high 
redshift may be systematically brighter, rarer, more biased objects than those at low redshift, so
the bias can be expected to evolve in a way that will be difficult to reliably calibrate.   
One principle goal of this paper is to examine the extent to which this 
impacts the cross correlation method, particularly whether the systematic biases involved are 
substantial compared to the resolution and accuracy of the photometric distribution that is recovered. 

To be very specific, on linear scales the relationship between the cross correlation function
and the (observable) autocorrelation function of the spectroscopic sample is given in equation 
\ref{eqn:xiprop} (further corrections are required for translinear scales).
\begin{eqnarray}\label{eqn:xiprop}
\xi_{ps}(r)=\frac{b_p}{b_s}\xi_{ss}(r)
\end{eqnarray}
Thus we may write 
\begin{eqnarray}\label{eqn:wpsdef}
w_{ps}(\theta,z_i)=\int_0^\infty \frac{b_p(z)}{b_s(z)}\xi_{ss}(r(z,z_i,\theta))\,\phi_p(z) \, dz
\end{eqnarray}
Since $b_s(z)$ can be fit with the spectroscopic data, this means that we can only invert the relation 
to solve for the product $b_p(z)\phi_p(z)$ in terms of observable quantities.  
This degeneracy cannot
be resolved by measuring the angular correlation function of the photometric sample.  
We can express $w_{pp}$ in terms of $\xi_{ss}$
\begin{eqnarray}
&w_{pp}(\theta)= \hspace{7cm} \nonumber \\
&\int dz_1 \int dz_2\, \phi_p(z_1) \phi_p(z_2)\, 
   \left(\frac{b_p(z_1) b_p(z_2)}{b_s(z_1) b_s(z_2)} \right) \xi_{ss}(\theta,z_1,z_2)
\end{eqnarray}
Changing variables to a central redshift and a $\Delta z=z_1-z_2$, and taking note 
that $\xi_{ss}$ vanishes for large $\Delta z$, we find that 
\begin{eqnarray}
w_{pp}(\theta) \propto \int dz \, \phi_p^2(z) \left( \frac{b_p^2(z)}{b_s^2(z)}\right) \xi_{ss}(\theta,z)
\end{eqnarray}
In terms of known observables, this relation can be inverted to determine the product 
$b_p^2(z)\phi_p^2(z)$.  Unfortunately, this quantity has the same direction of degeneracy as the 
product $b_p(z)\phi_p(z)$.  Thus the observable $w_{pp}(\theta)$ can only be used to improve the 
accuracy to which the product is determined, it cannot be used to break the 
degeneracy between the large scale bias and the selection function.   In order to obtain $\phi_p(z)$ 
it will be necessary either to appeal to some model of the bias evolution 
or to find another observable that can be used to break the degeneracy. 

This is significant because estimators of e.g. the mean redshift of a sample will be affected by 
assuming a functional form for the bias that is incorrect.   If we know we are recovering the product
$b(z)\phi(z)$ then in our estimator of the mean redshift
\begin{eqnarray}
\bar{z}_{\rm est}=\int_0^\infty z\, \frac{b_{\rm true}(z)}{b_{\rm est}(z)}\, \phi(z) \, dz
\end{eqnarray}
while the true $\bar{z}$ is 
\begin{eqnarray}
\bar{z}_{\rm true}=\int_0^\infty z\, \phi(z)\, dz 
\end{eqnarray}
It is not unreasonable to suspect that the bias may not be a particularly smooth in its transitions if 
the sample of galaxies accessible to a photometric survey shifts abruptly as some redshift threshold
is crossed.  One example of a very rapidly changing bias function can be seen in table 2 of  \cite{2009MNRAS.397.1862P} 
where the bias of a sample of LRG galaxies is computed in several photometric bins.  At a redshift
of $z\sim0.35$, the bias jumps dramatically from 1.77 to 2.36, and back down again (somwhat more 
smoothly) to 1.9 at higher $z$.  
As a quick illustrative example, we take the two functional forms of $\phi_p(z)$ used later in this paper 
(see sec. \ref{sec:mocks}) and compute the error in $\bar{z}$ that occurs if we assume a smooth transition in the galaxy bias from 1.7 to 1.9 in $b_{\rm est}$, but allow the true bias $b_{\rm true}$ 
to jump to 2.3 in between.  We find that the fractional error in the 
mean redshift is 6\% and 11\% if the interval is $0<z<1$ and the jump is placed at $z=0.8$.  If the jump
is place at $z=0.3$, the fractional error is 0.05\% and 0.2\%, which is somewhat less significant, since 
there is substantially less volume at $z=0.3$.
\section{Tests with mock data}\label{sec:sims}
\subsection{Simulations and mock galaxy catalogs}\label{sec:mocks}
We use the halo model to populate N-body cold dark matter simulations with mock 
galaxies to test the cross-correlation method.  The simulations compute the 
evolution of large scale structure in a periodic, cubical 
box of side $1$ Gpc$/h$ using a Tree-PM code \cite{2002ApJS..143..241W}.  
There are $1024^3$ dark matter particles of mass $6 \times 10^{10} M_\odot/h$.
The randomly generated Gaussian initial conditions are evolved from a starting
redshift of $z=75$.  The Plummer softening is $35$ kpc$/h$ (comoving).  

Halo catalogs are constructed from this simulation using a Friends of Friends algorithm 
\cite{1985ApJ...292..371D} with a linking length of b = 0.168 in units of the mean 
inter-particle spacing.  There are approximately 7.5 million halos of mass greater than 
$5 \times 10^{11} M_\odot/h$ resolved with $\sim 8$ or more dark matter particles.  The galaxy 
catalogs are constructed by populating these halos according to the halo-model prescription.
It is assumed that very small halos host no galaxies.  Halos that cross a mass threshold
$M_{\rm min}$ are assumed to host one central galaxy.  Halos of much higher mass are assumed
to have formed through mergers of smaller halos and will host both
central and satellite objects.  The mean number of galaxies in a halo of mass $M$ is given by 
\begin{eqnarray}\label{eqn:hod}
\left<N_{\rm gal}(M)\right>=\Theta(M-M_{\rm min}) 
\left(1+\frac{M-M_{\rm min}}{10\,M_{\rm min}}\right)
\end{eqnarray}
The position of the central galaxy is assumed to be at the halo center defined by the position
of the most 
gravitationally bound dark matter particle.  The number of satellite galaxies
in any particular halo is drawn from a Poisson distribution, and the satellites are assumed 
to trace the dark matter.

We use this prescription to populate the simulation with several different galaxy samples.  One population, 
used as the mock photometric catalog, has a relatively low value of $M_{\rm min}=5\times10^{11} M_\odot/h$;
these are fairly common galaxies with low value of the large-scale bias $b_p$ 
with respect to the dark matter.  Although we know the true value of the redshifts of these objects from the 
simulations, we do not use any redshift information for the photometric sample when testing the reconstruction 
algorithm developed here.  
The other populations we have created are mock spectroscopic samples, with values of $M_{\rm min}=
1\times10^{12} M_\odot/h$ and $ 7\times 10^{12} M_\odot/h$ that generate much rarer mock galaxies with higher 
biases $b_s$ with respect to the underlying dark matter.  For some tests of the cross-correlation method, we also
use a fair subsample of  50\% or 80\% of the mock photometric population to define the spectroscopic sample.

\begin{figure}
\begin{center}
\resizebox{3.5 in}{!}{\includegraphics{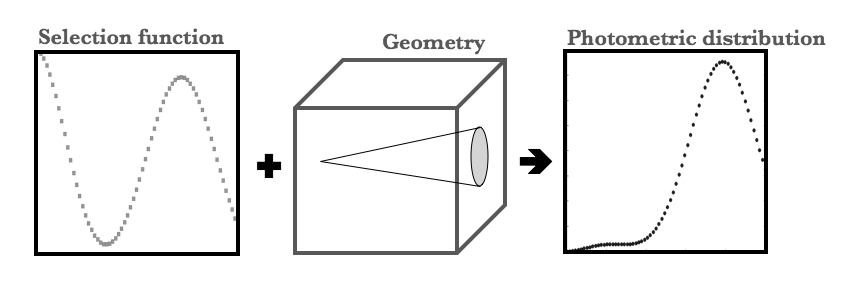}}
\end{center}
\caption{The final photometric redshift distribution is affected by the selection function and also the survey geometry.}
\label{fig:phicartoon}
\end{figure}

To test the cross-correlation method, we also need to impose both a selection function and an observation window 
on the photometric sample.  For convenience, we choose to perform the reconstruction in bins of comoving distance $\chi_i$ rather than redshift bins $z_i$, 
but the method can be used with either choice of bins.
We place the observer in the center of one face of our simulation cube, and assume that 
(s)he observes a conical volume with a 12 degree opening angle.  The cone stretches the length of the box ($1$ Gpc$/h$)
along the line of sight.  We adopt a toy model for the selection function, the sum of two Gausseans, which defines
the fraction of galaxies ``detected"  $N_{\rm keep}(\chi)/N(\chi)$ at each of $70$ slices in comoving
distance $\chi$ through the box. 
\begin{eqnarray}\label{eqn:sf}
\frac{N_{\rm keep}(\chi)}{N(\chi)}\propto  \hspace{5.5 cm} \nonumber \\
\frac{1}{\sqrt{2\pi\sigma_1^2}} \, e^{-(\chi-\chi_1)^2/2\sigma_1^2}
+ \frac{1}{\sqrt{2\pi\sigma_2^2}} \, e^{-(\chi-\chi_2)^2/2\sigma_2^2}
\end{eqnarray}
We have normalized this quantity so that the maximum value is 1, and we refer to it as the ``detection 
fraction" later in the paper.  We present two models for comparison in this paper, 
$[\chi_1, \sigma_1,\chi_2, \sigma_2]=[0,0.15,0.8,0.16]$ and $[0.3,0.07,0.7,0.10]$.
In the limit of small galaxy numbers in the slice,
it is significant that we apply the selection function before the geometry for reasons of cosmic variance.
The shape of the final photometric redshift distribution is affected both by the selection function and 
the geometry of of the mock observation.  This is illustrated graphically in figure \ref{fig:phicartoon}, which shows the first of the two selection functions.  
Once we have mock photometric and spectroscopic samples in place, we compute in each bin $\chi_i$ 
the autocorrelation function of the spectroscopic sample $\xi_{ss}(r,\chi_i)$, and the angular cross correlation between the spectro-z objects in the $i^{th}$ bin and the entire photometric sample 
$w_{ps}(\theta,\chi_i)$.  We have used the algorithm of Landy and Szalay 
\cite{1993ApJ...412...64L} to compute the
correlation functions.  To compute $w_{ps}(\theta,\chi_i)$ we opt to measure the 3D correlation function 
$\xi_{ps}(r,\chi_i)$ and perform the integral of eqn. \ref{eqn:wpsdef} numerically, because the mock observation 
volume is small which renders direct calculation of $w_{ps}(\theta,\chi_i)$ noisy.  
This will not be a problem for surveys that cover a large fraction of the sky. 

We are eager to test how sensitive the reconstruction of the photometric distribution 
function $\phi_p(\chi)$ may be to evolution of the bias $b_p$ that is not accounted for.  We
addressed this question analytically in section \ref{sec:theory}, but it is useful to examine the 
issue with simulations to see if the systematic
biases that occur are significant compared to the error bars on the re-constructed distribution. 
Evolution of the bias can occur because of evolution in the underlying large scale structure, and
it can occur because the population of photometric
objects detected by an instrument evolves as a function of redshift, as  
expected in a magnitude limited sample.  Our simulation volume is made of a single time-slice, so
we cannot examine the effects of the first mechanism at present, but we expect the effects of the 
second mechanism to dominate the bias evolution. We have devised a simple method inspired
by the results of \cite{2006MNRAS.371.1173V} and \cite{2009ApJ...696..620C} 
to introduce this type of bias evolution into our 
mock galaxy sample.   Whereas before we applied the selection function in eqn \ref{eqn:sf} randomly
to the galaxies in each slice, now we choose to keep galaxies preferentially that live in the largest
halos.  We assume that the brightest galaxies live in the biggest halos and that central galaxies 
are brighter than satellite galaxies at a given redshift.  
First we rank order the halos in the slice.  We determine the number of galaxies to 
be kept from eqn. \ref{eqn:sf}, and place one in the center of each halo beginning with 
the halo of highest mass.  If the number of galaxies exceeds the number of halos in the slice, 
we begin placing satellite galaxies.  We again start with the largest halo and continue populating 
each halo with satellites until we have placed all the galaxies.  This procedure generates a mock 
catalog of photometric galaxies whose clustering strength will depend strongly on the selection 
function.  The effect of this procedure on the largescale bias is somewhat complicated.  In the regime
where only satellites in the lowest mass halos are being eliminated, if the detection fraction from eqn. \ref{eqn:sf} is high the bias will be close to the bias of the whole population.  However as the detection
fraction gets lower, the sample will be more highly biased with respect to the dark matter because only 
satellites in the largest halos are being kept, thus weighting the largest halos more heavily than the smaller halos in the clustering measurement.  However, if the detection fraction is so low that all of the satellites are eliminated and the population consists only of centrals, the largescale bias will be lower than for the whole population, because more weight from the largest halos has been discarded than from the smaller halos.   In this analysis we are principally in the second regime, which means the 
large scale bias will tend to follow the shape of the selection function.  

\begin{figure}
\begin{center}
\resizebox{3.7 in}{!}{\includegraphics{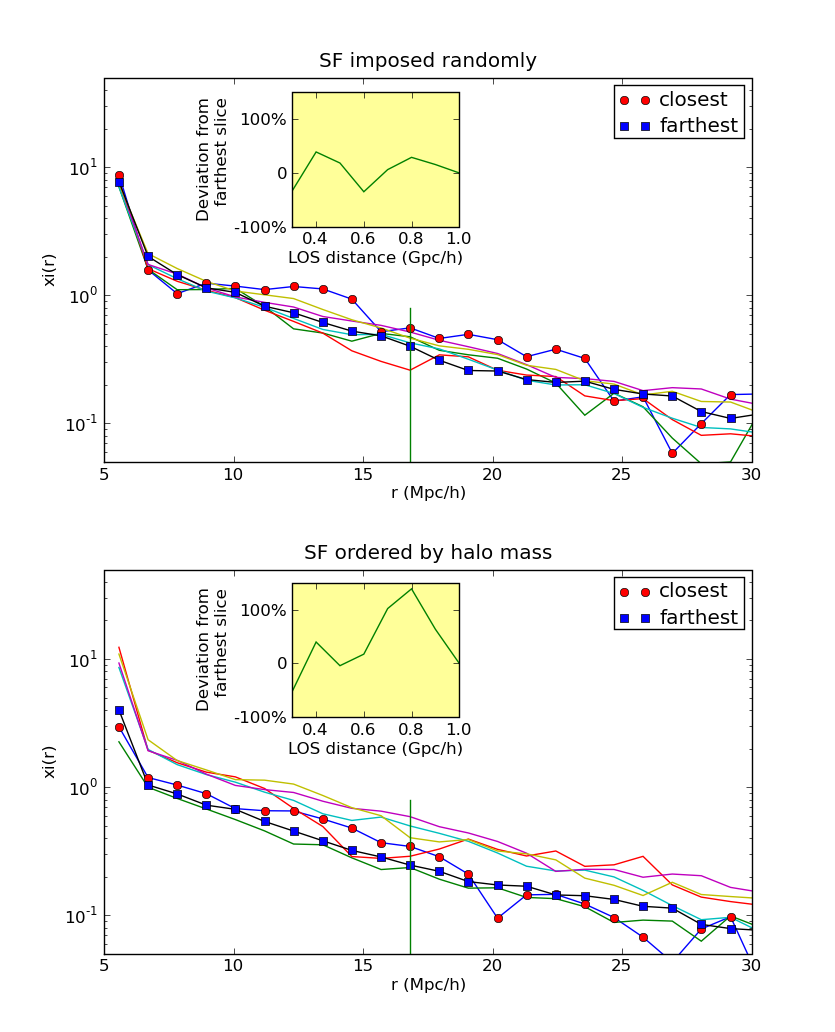}}
\end{center}
\caption{The 3D correlation function of the photometric sample in each slice along the line of sight.  The 
closest and farthest slice plotted are labeled with symbols.  The inset shows the relative variation among 
the curves at a single scale.   On the top the there is much less variation among the curves than on the bottom.  The inset on the bottom clearly reflects the shape of the 
selection function (and hence the bias) used to create the sample.  }
\label{fig:bevol}
\end{figure}

This is demonstrated in fig. \ref{fig:bevol}, which plots the correlation function measurement of the mock photometric sample
in each conical slice.  The top panel shows the result if the galaxies are eliminated randomly, and the bottom panel shows the results from the procedure we just described.  The inset shows the variation in the value of $\xi_{pp}(r)$ along the line of sight (where we have picked a particular scale, indicated with a vertical line).  On the top we show that compared to the last slice (marked with blue squares), the value of $\xi_{pp}(r)$ varies by some tens of percent, and is a relatively flat function of the line-of-sight distance.  
The bottom inset on the
other hand, clearly reflects the shape of the redshift distribution function used to create the sample (plotted later in the bottom panel of fig. \ref{fig:realist}) , and shows variations in the normalization of 
$\xi_{pp}(r)$ of up to 150\%. 

We emphasize that
we do not expect this method to quantitatively capture the bias evolution in a real galaxy survey, but
we expect it is qualitatively similar, and as long as the bias evolves significantly, 
it will be useful in testing the magnitude of the effect
on the reconstruction.  We will refer to this method of applying the selection
function as the Ordered Selection Function (OSF), and the simple random method as the Random 
Selection Function (RSF).  By comparing the reconstruction in the two cases, we determine how 
important it is to account for evolution in the bias.  

\subsection{Pipeline}
Once the cross correlation $w_{ps}(\theta,\chi_i)$ and autocorrelation function $\xi_{ss}(r,\chi_i)$ 
have been measured, a procedure is needed to perform the inversion of the integral of eqn.
\ref{eqn:wpsdef} to obtain the line of sight distribution of the photometric sample in some bins $\chi_j$,
and to obtain error bars on those measurements.  For a single angular scale $\theta$ and
bin $i$ in spectroscopic redshift we approximate the 
integral in eqn. \ref{eqn:obs} as a discreet sum over $N$ redshift bins (ignoring bias evolution
for the moment).
\begin{eqnarray}\label{eqn:wpssum}
w_{ps}(\theta,\chi_i)=  
\int_0^\infty \xi_{ss}(r(\chi,\chi_i,\theta))\,\phi_p(\chi) \, d\chi \nonumber \\
\approx \sum_{j=0}^{N-1} X_{ij} \phi_{p,j}  \label{eqn:ximat}
\end{eqnarray}
\begin{eqnarray}
&X_{ij}=\frac{b_p}{b_s}\xi_{ss}\left(\chi_i,r=\sqrt{\left(\theta \chi_i \right)^2+\left(\chi_j-\chi_i\right)^2}\right)
\end{eqnarray}
The observation will be performed in multiple angular bins $\theta$. 
The values of $r$ in $X_{ij}$ will not be at the exact positions where we have made the measurement
of $\xi_{ss}$ (in our case in 60 bins between 5 and 50 comoving Mpc/h).  
If the $r$ falls within the domain of our data, 
we interpolate $\xi_{ss}$ with a spline, if it is larger than the biggest scale we measure, 
the value of $\xi_{ss}$ is extrapolated using 
a fit of the form $r^{-2}$ to a subset of data points (larger than 15 Mpc/h) 
measured in the volume.  If the value of $r$ is less than
the minumum value of $r$ measured, we reject the entire $\theta$ bin, 
since we do not trust the method for $\xi_{ss}$ measured on very small scales.   

Collecting all the observed 
$w_{ps}$ values in a single vector ${\bm w}$ (one element
for each unique combination of $\chi_i$ and $\theta$), the solution for $\phi_p(\chi_j)$ will be 
obtained by inverting relation
\begin{eqnarray}\label{eqn:linear}
{\bm w}={\bf X} \cdot {\bm \phi}
\end{eqnarray}
Here, ${\bf X}$ is a matrix whose elements are given by eqn. \ref{eqn:ximat}, and whose dimension
is 
\begin{eqnarray}
{\rm (\# spec \, bins\,} i * {\rm \# theta\,bins)} \times N \nonumber
\end{eqnarray}  $\bm \phi$ will 
be a vector of length  $N$,and $\bm w$ will be a vector of length 
(\# spec bins $i *$ \# theta bins).  It is significant that measurements at different values of 
$\theta$ can mix in the inversion; since they are correlated it is important that they do so. 
We do not know the pre-factor $b_p/b_s$, but as long as it is assumed to be constant
(as in the special case where the spectroscopic sample is a {\it fair} subsample of the
photometric population) this is not relevant.  The reconstructed $\phi_p(\chi_j)$ will not be correctly normalized after the matrix inversion because there is no information 
in eqn.  \ref{eqn:wpssum} about the 
total number of galaxies in the photometric sample.  
The normalization of $\phi_p(\chi_j)$ will be set by requiring that it integrate to 1.  

   In principle, solving equation \ref{eqn:linear} for $\phi_p(\chi_j)$ is a simple matter of inverting a 
matrix, but in practice doing so is numerically unstable and furthermore 
errors in the observables ${\bm w}$ and ${\bf X}$ must be accounted for 
in the solution.  Since ${\bm w}$ is a projection through the same dark 
matter structures traced by ${\bf X}$, there will be non-negligible correlations between the errors in the
two observables.   To extract $\phi_p(\chi)$, we write down the expression for the $\chi^2$ statistic,
and minimize it.  Typically we are attempting to reconstruct $\phi_p(\chi_j)$ in as many bins $N$ as
possible.   There is often insufficient information in the correlation 
functions to completely constrain $\phi_p$ in the chosen number of bins, thus it becomes necessary to stabilize against solutions that have highly anti-correlated adjacent bins.  Since it is reasonable to 
assume that the true distribution will not be highly oscillatory, we adopt a smoothness prior to 
regularize the solution, and add it to our expression for $\chi^2$ below.  
\begin{eqnarray}\label{eqn:chisq}
\chi^2=({\bm w}-{\bf X}{\bm \phi})^T {\bf C}^{-1}_{\bm \phi}({\bm w}-{\bf X}{\bm \phi})
+ \lambda \,{\bm \phi}^T {\bm B}^T {\bm B} {\bm \phi}
\end{eqnarray}
Here ${\bf C}_{\bm \phi}$ is the covariance matrix incorporating errors in both ${\bm w}$ 
and ${\bf X}$, and the subscript denotes that it is an
explicit function of ${\bm \phi}$.  For purposes of the present analysis, however, we will
assume that the spectroscopic correlation function is perfectly determined, and 
we will propagate errors from $w_{ps}$ only, specifically ${\bf C}^{-1}_{\bm \phi}={\bf C}^{-1}=$
inverse covariance matrix of $w_{ps}(\theta,\chi_i)$.  This is not a good approximation as we 
will soon show, and we refer the reader to appendix \ref{app:error} for a description of the 
full error propagation from ${\bf C}^{-1}_{\bm \phi}$ 
into errors on $\phi_p(\chi_j)$.  The reader should consider all errors 
reported on $\phi_p(\chi_j)$ as lower limits on the total error.

The regularization scheme we have adopted is described in detail in \cite{2002nrc..book.....P} 
section 18.5.  
The intention is to add a term to $\chi^2$ that gets large when neighboring points have widely
different values.  Minimization of $\chi^2$ will then tend to solutions that do not have anti-correlated neighboring points.  The matrix ${\bm B}$ is given by
\begin{eqnarray}
{\bm B}=\left( {\begin{array}{ccccccccc}
-1 & 1 & 0 & 0 & 0 & 0 & 0 & \cdots & 0  \\
 0 & -1 & 1 & 0 & 0 & 0 & 0 & \cdots & 0  \\
\vdots  & & & & \ddots & & &  & \vdots  \\
 0 &\cdots & 0 & 0 & 0 & 0 & -1 & 1 & 0  \\
  0 & \cdots & 0 & 0 & 0 & 0 & 0 & -1 & 1  \\
 \end{array} } \right)
\end{eqnarray}
Note that ${\bm B}$ has one fewer row than column.  The factor $\lambda$ should be 
chosen such that the first and second terms in $\chi^2$ contribute roughly equal weight. 
This can be approximately arranged if $\lambda$ is taken to be 
\begin{eqnarray}
\lambda=\frac{{\rm Tr}({\bm X}^T {\bm C}^{-1}{\bm X})}{{\rm Tr}({\bm B}^T{\bm B})}
\end{eqnarray}
but in practice, the weight in $\chi^2$ will also depend on the solution and the values
of $w_{ps}$.  We find that the
quality of the reconstruction depends on how well the two terms in eqn. \ref{eqn:chisq} are
balanced.  Since this depends on the answer, we opt to refine the value of lambda
and re-compute the solution iteratively as follows.  
\begin{itemize}
\item Compute a tolerance parameter defined from the two terms in $\chi^2$ as tol=1-(term 1/term 2).  
\item If tol $>0$, $\lambda$ is  
increased by a factor of 10.  If tol $<0$ we decrease it by 10.  
\item We recompute the solution and the tol
\item If the absolute value of tol has decreased, we repeat the refinement of lambda, otherwise 
we exit and keep the previous value of the solution
\end{itemize}
In practice, the procedure usually requires only 1 refinement of $\lambda$.  We observe that 
changing the algorithm to have smaller steps in lambda (e.g. a factor of 2 rather than 10) does not 
improve the solution, and occasionally over-smooths it.   We emphasize that while we have identified 
an algorithm that works, we have not optimized the application of a smoothness prior.  Since the 
solution is moderately sensitive to the smoothing, care should be taken to understand the properties 
of the smoothing before applying this technique to real data.  We have not studied the 
effects of different smoothing algorithms 
because it is likely that the need for a smoothness prior will be eliminated
in future analyses by using the photometric probability distributions as a 
prior instead.  Since we have no mock photometric redshift probabilities, examining that technique 
is beyond the scope of this paper, but will be the subject of further research. 

To minimize $\chi^2$ we take the derivative with respect to ${\bm \phi}$ and set it equal to zero.
Taking ${\bm C}^{-1}$ to be independent of ${\bm \phi}$ and noting that it is symmetric we find
\begin{eqnarray}
-2{\bm w}^T {\bm C}^{-1} {\bm X} + 2 {\bm \phi}^T {\bm X}^T {\bm C}^{-1} {\bm X} +2\lambda
{\bm \phi}^T{\bm B}^T{\bm B} = 0 \hspace{0.5cm}  \label{eqn:minchi2}
\end{eqnarray}
\begin{eqnarray}
{\bm X}^T {\bm C}^{-1}{\bm w} = \left[{\bm X}^T {\bm C}^{-1} {\bm X} +\lambda
{\bm B}^T{\bm B}\right] {\bm \phi} \\ \label{eqn:answer}
 {\bm \phi}= \left[{\bm X}^T {\bm C}^{-1} {\bm X} +\lambda
{\bm B}^T{\bm B}\right]^{-1} {\bm X}^T {\bm C}^{-1}{\bm w}
\end{eqnarray}
The covariance matrix of the recovered ${\bm \phi}$ is given by 
\begin{eqnarray}\label{eqn:phierrs}
{\rm Cov}[\bm \phi] \propto \left[{\bm X}^T {\bm C}^{-1} {\bm X}+\lambda
{\bm B}^T{\bm B}\right]^{-1}
\end{eqnarray}

To recover the constant of proportionality, we will need to rescale the elements of the 
covariance matrix by the square of the factor used to rescale $\bm \phi$ (since we 
renormalize such that $\phi(\chi)$ integrates to 1).   We caution that all matrix multiplications above
are finite sum approximations to integral quantities, so care must be taken to ensure that phi and its 
error bars come out to scale.  For clarity, let us refer to $w_{ps}(z_i)$ as 
a function of a continuous variable $z_i$ (which will label the spectroscopic bins).  $\xi(z_i,\chi_i)$
will be a function of both $z_i$ and another continuous variable $\chi_i$ (which will label the 
bins in which $\phi$ is reconstructed).   We are considering a single value of $\theta$, and $z$ and 
$\chi$ can represent either redshifts or comoving distances, we simply use both 
letters to distinguish them.  Ignoring regularization, the continuous expression for $\chi^2$ is then
\begin{eqnarray}\label{eqn:contin}
\chi^2=\int dz_1 dz_2\left[
\left(w_{ps}(z_1) - \int d\chi_1 \phi(\chi_1) \xi(z_1,\chi_1) \right) \right. \nonumber \\
\left. C^{-1}(z_1,z_2)
\left(w_{ps}(z_2) - \int d\chi_2 \phi(\chi_2) \xi(z_2,\chi_2) \right) \right] \hspace{0.2cm}
\end{eqnarray}
To minimize $\chi^2$ we must set the functional derivative $\delta \chi^2/\delta \phi(\chi)=0$.  We
leave the details to the reader, but note that in eqn \ref{eqn:minchi2}, the factors of $\Delta z$ that
correspond to integrals over $z$ cancel but the factor of $\Delta\chi$ does not.  
\subsection{Redshift Distribution Reconstruction}
\begin{figure}
\begin{center}
\resizebox{3.7 in}{!}{\includegraphics{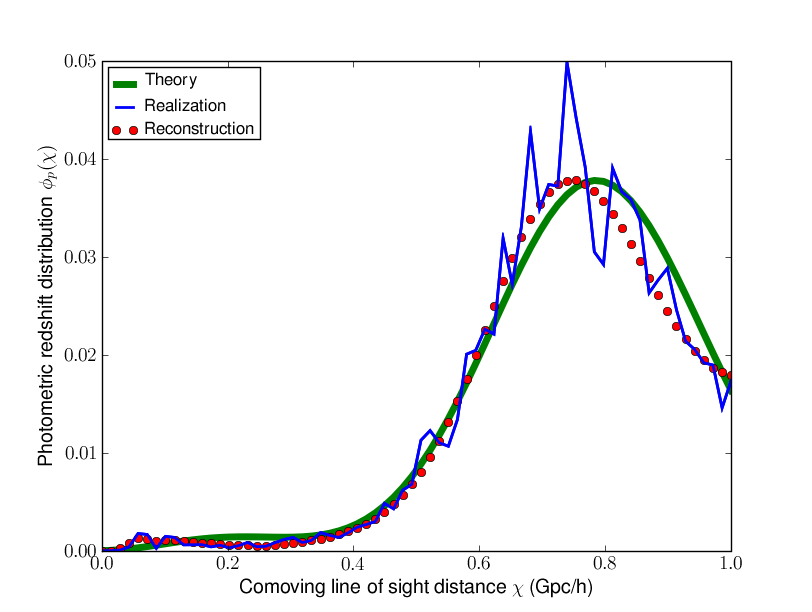}}
\end{center}
\caption{The photometric distribution $\phi_p(\chi)$ and its reconstruction.  
The x axis is comoving line of sight form the observer, who is located at the origin.
The thick solid line shows the theoretical value of the distribution, 
the thin wavy line shows 
the particular realization in this simulation volume, and the points show the reconstructed solution.}
\label{fig:one}
\end{figure}

In this section we present a series of reconstructions that examine various aspects of the problem
and identify the potential difficulties in applying this method.  We begin in fig. \ref{fig:one} with the simplest case and gradually add complexity. The heavy solid line shows 
the theoretical redshift distribution (selection function + geometry) that has been applied mock photometric catalog of galaxies with $M_{\rm min}=5 \times 10^{11}$ in eqn. \ref{eqn:hod} .    This 
redshift distribution corresponds to the selection function shown on the left of fig. \ref{fig:phicartoon}. 
We have chosen galaxies randomly (the RSF method described in
section \ref{sec:mocks}) in applying the selection function.  The resulting catalog has 54,000 galaxies.  We have measured the detection fraction in each slice for this particular realization, and plot it with a thin wavy line in fig. \ref{fig:one}.   The spectroscopic calibrating sample is taken to be a different population of galaxies with $M_{\rm min}=7 \times 10^{12}$ yeilding around 5600 galaxies, 
i.e. a highly biased tracer of the density field compared
to the photometric sample.   We assume both the observables are measured with perfect accuracy.
Since we have assumed perfect knowledge of the spectroscopic 
correlation function, we opt to measure it in the entire light cone, and us the result in each of the $i$ spectroscopic bins $\xi_{ss}(r,\chi_i)=\xi_{\rm whole \, volume}(r)$.  This is only justified because there is no evolution in the light 
cone, and the calibrators are not affected by the selection function which changes along the line of sight.  
We perform the reconstruction by computing the matrices 
${\bm X}$, ${\bm w}$, ${\bm C}^{-1}$, and ${\bm B}$ and using them in eqn. \ref{eqn:answer}.  
The result is plotted in fig. \ref{fig:one} with points.   Notice that the reconstruction follows the values of this
realization rather than the theoretical redshift distribution used to create the sample. 

\begin{figure}
\begin{center}
\resizebox{3.7 in}{!}{\includegraphics{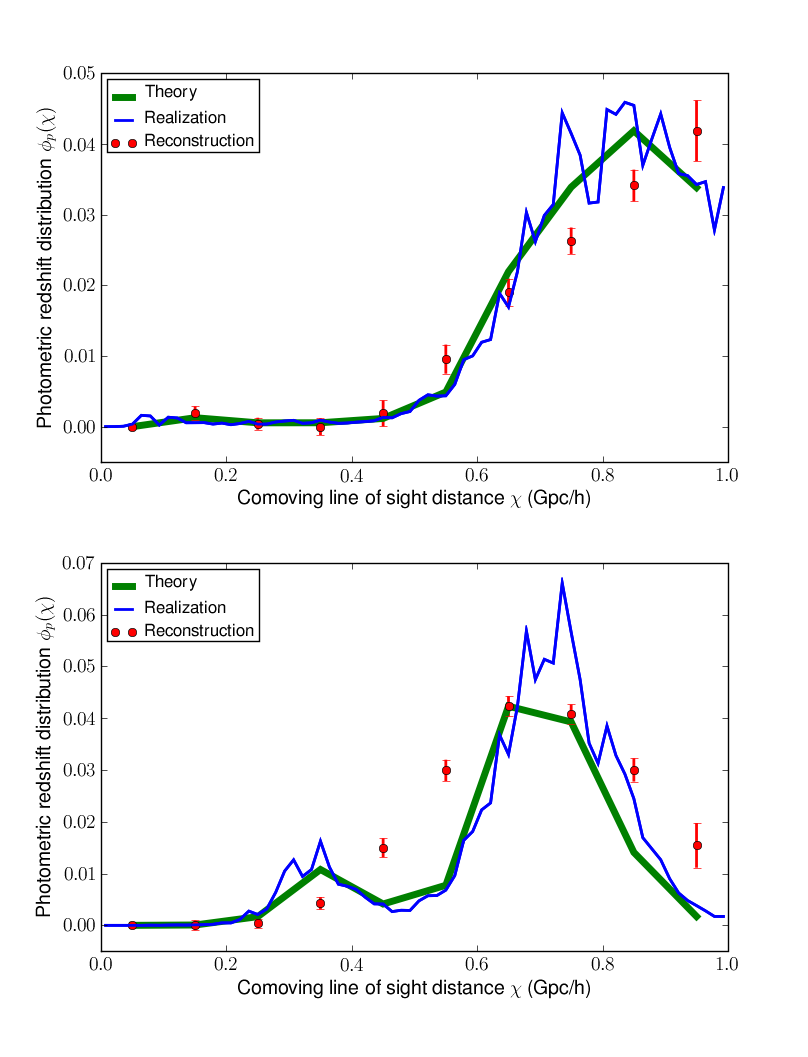}}
\end{center}
\caption{The photometric distribution $\phi_p(\chi)$ and its reconstruction.  
The top and bottom panels show two different 
selection function choices.  Each is the sum of two Gaussians with 
$[\chi_1,\sigma_1,\chi_2,\sigma_2] = [0.0,0.15,0.8,0.16]$ on the top and $[0.3,0.07,0.7,0.10]$ 
on the bottom.  The errors come from Poisson error in the cross-correlation measurement.  The spectroscopic sample is made of rare objects with $M_{\rm min}=7 \times 10^{12}$
in eqn. \ref{eqn:hod}. There are very few objects in each conic section so the correlation functions are 
poorly measured, and this reconstruction is not capturing the bimodal behavior.}
\label{fig:realist}
\end{figure}

Fig. \ref{fig:realist} reveals a more realistic picture.  The lines and points in fig. \ref{fig:realist} are all the same as in fig. \ref{fig:one}.  We now measure the autocorrelation of the 
spectroscopic sample $\xi_{ss}(r,\chi_i)$ in each of the conic sections, and use those measurements
to perform the reconstruction.  The top and bottom
panels show two different choices of selection function, whose parameters are marked in the caption. 
The top is difficult to reconstruct 
because it peaks at $\chi=0$ where the is no volume in the mock observation, and the bottom is 
a challenge because the feature coincides with the bin spacing, which will be a problem to 
reconstruct because of the smoothness prior.  We have drastically decreased the number of 
bins, so that there are a reasonable number of calibrators in most of the bins.  The normalization criterion, which comes from the condition that $\phi_p$ integrate to 1, becomes more 
sensitive to noise in the reconstruction when the number of bins is decreased.  In this plot and the plots that follow we set the normalization by hand so that we can illustrate other points; we let the maximum
value of the reconstruction (points) equal to maximum value of the theoretical input $\phi$ (thick solid
line).   While the reconstructions in fig. \ref{fig:realist} show the correct trends, they are not sufficiently high quality to recover the bimodal behavior in the reconstructed selection function.

We now briefly discuss the error bars in fig.  \ref{fig:realist} obtained from eqn. \ref{eqn:phierrs}.  
The ${\rm Cov}[\phi]$ matrix is not diagonal; we conservatively report $1/\sqrt{{\rm Cov}^{-1}[\phi]_{i,i}}$ as the error
on the $i^{th}$ bin.  These error bars represent a lower bound on the error because we have only 
propagated error from $w_{ps}$ and not from $\xi_{ss}$. 
The matrix ${\bm C}^{-1}$ in eqn. \ref{eqn:phierrs} is the inverse
covariance matrix of the cross-correlation measurement ${\bm w}$, and with sufficient volume can be estimated by dividing the observation into a number of bins and bootstrapping.  Here, however, we opt to follow \cite{1980lssu.book.....P} 
and assume that on these scales the errors are dominated by Poisson noise.  In this case, ${\bm C}^{-1}$ is diagonal, and its elements are given by 
\begin{eqnarray}
{\bm C}_{ii}=\frac{1}{\sqrt{n_{\rm pairs}}}=\bar{n}_s \bar{\Sigma}_p \chi_s^2 \Delta\chi_s \, \Omega \sin(\theta)\Delta\theta 
\end{eqnarray}
We use survey parameters appropriate to large upcoming missions.  We take $\bar{n}_s=1\times 10^5$ galaxies per  $(Gpc/h)^3$, 
$\bar{\Sigma}_p=100$ photometric galaxies per square arcmin, and $\Omega=20,000$ square 
degrees on the sky.   $\Delta\chi_s$ is the width of the bin of spectroscopic data.  The resulting error bars in  \ref{fig:realist} and those that follow display a surprising
trend.  Even though the number of pairs is dramatically larger for bins at farther comoving distance, the 
reconstruction of the photometric distribution is less accurate in the most distant bins.  The key to 
understanding this trend is that for a fixed angular scale $\theta$, the physical scale being probed in 
distant bins is much larger than at nearby distances.  Since the correlation function is a steeply 
dropping function of separation, the impact of shrinking values in $\bm X$ of eqn. \ref{eqn:phierrs} 
dominates over the growing number of pairs in $\bm C^{-1}$.  By measuring $w_{ps}$ on smaller angular scales the errors in the farthest bins can be improved, but there is a limit to how far this can be pushed because measurements at small angular scales will send the near field into the trans-linear and 1-halo regime, 
for which there are corrections to the underlying assumption that $\xi_{ps} \propto \xi_{ss}$.

\begin{figure}
\begin{center}
\resizebox{3.7 in}{!}{\includegraphics{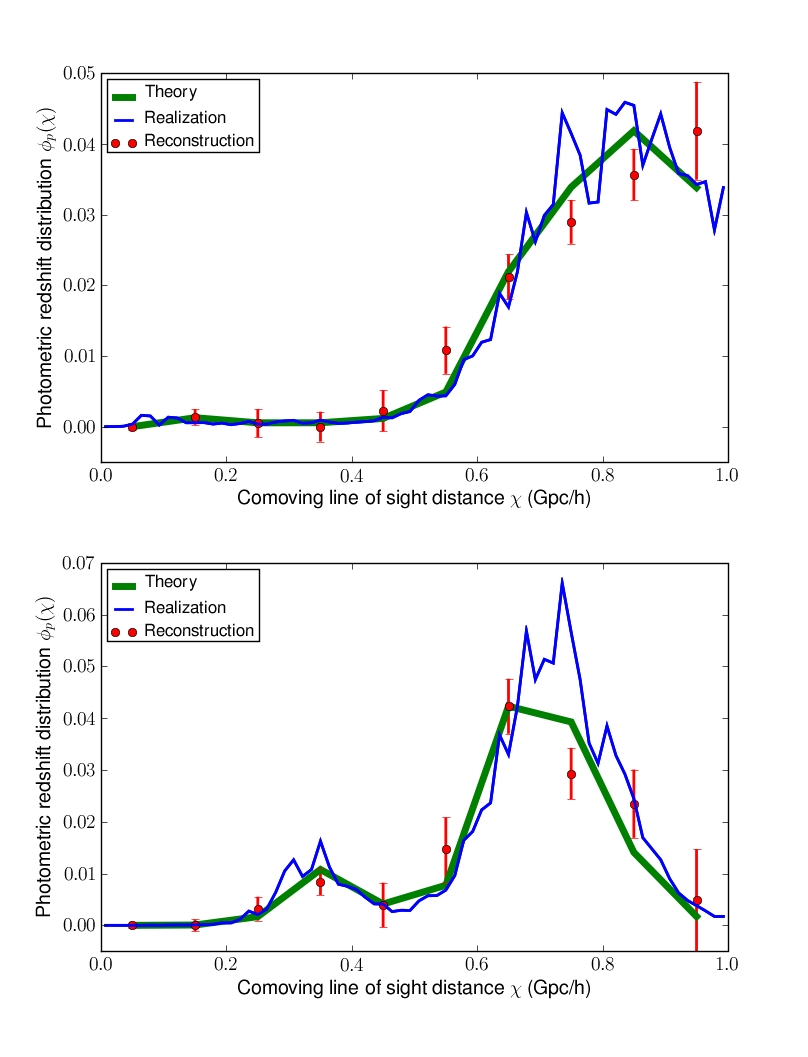}}
\end{center}
\caption{The photometric distribution $\phi_p(\chi)$ and its reconstruction.  
Now the spectroscopic population has $M_{\rm min}=1 \times 10^{12}$ and is roughly the 
same size as the photometric one though it is more biased and comprised of different galaxies.  Because the correlation function has been measured well, the reconstruction recaptures the bimodal
behavior.}
\label{fig:three}
\end{figure}

In figure \ref{fig:three} 
we switch to a larger calibration sample with $M_{min}=1 \times 10^{12}$.  We see that the 
reconstruction now captures the bimodal behavior, but both the resolution (bin spacing)
and errors are modest and the smoothing is still evident in the last bin.  Notice that the error bars 
have increased: this is because this calibration sample has lower bias, so the elements in $\bm X$
in the covariance matrix of $\phi(\chi)$ (eqn. \ref{eqn:phierrs}) are all smaller.  This is an illusion 
however.  Had we propagated 
the error from $\xi_{ss}$, it would contribute larger errors to 
figure \ref{fig:realist} than to \ref{fig:three} because the measurment is noisier for the smaller sample. 
In moving to the larger sample, the number of spectra required has increased by
an order of magnitude, and is now the same size as the mock photometric catalog (after the 
selection function has been applied).  In summary, comparison of fig. \ref{fig:one}, fig. \ref{fig:realist}, and fig. \ref{fig:three} demonstrates that $\xi_{ss}(\chi)$ must be reasonably well determined for the 
shape of the reconstructed photometric distribution to be well captured.   It is worth noting 
that we have used no priors in the determination of $\xi_{ss}(\chi)$, improving and applying theoretical
priors may yield significantly better results with fewer spectra. 

The fact that the more numerous calibrators generated such improved results suggested to us that 
perhaps the reconstruction with the rarer calibrators could be improved by simply fitting the 
spectroscopic observations with a 2-parameter power law, and performing the reconstruction in that 
way.  This appears to discard too much information contained in $\xi_{ss}(\chi)$.  The result was visually similar to fig. \ref{fig:realist}, namely, the bimodal behavior in the distribution is lost. 

\begin{figure}
\begin{center}
\resizebox{3.7 in}{!}{\includegraphics{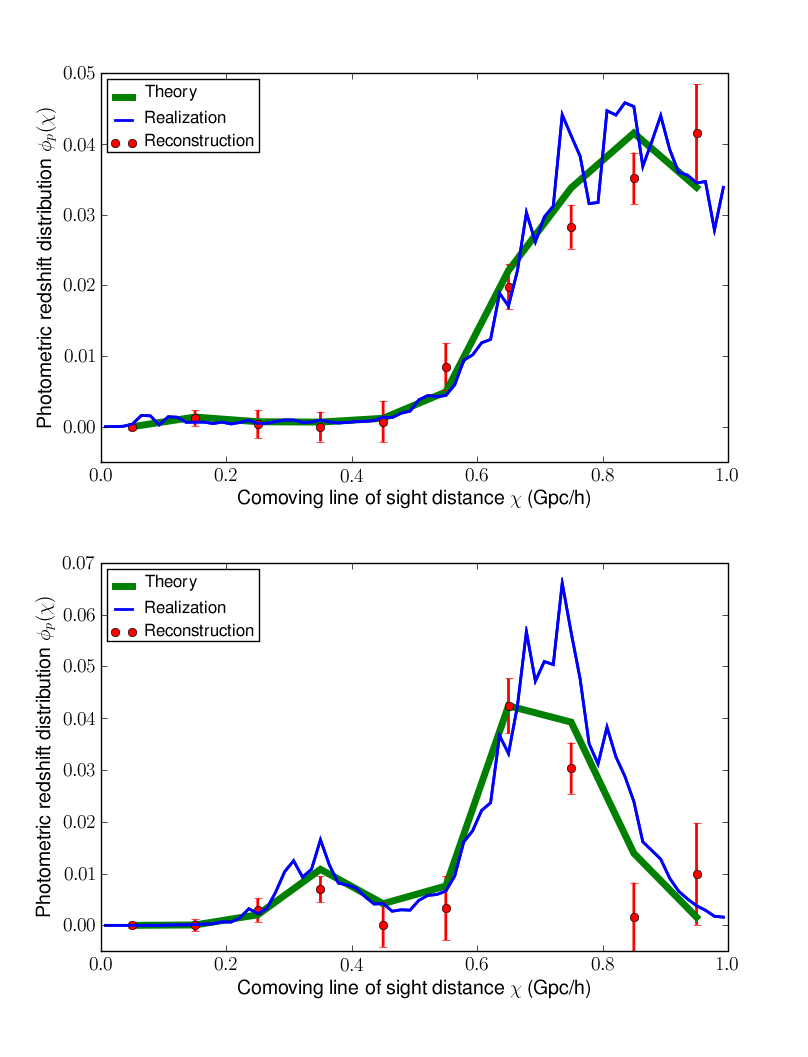}}
\end{center}
\caption{The photometric distribution $\phi_p(\chi)$ and its reconstruction.  Here the selection function 
applied to the photometric sample causes the bias to evolve as a function of comoving distance from 
the obsever.  The calibrators are the $M_{\rm min}=1 \times 10^{12}$ sample, whose bias does not evolve.  The result is a systematic shift of marginal significance in areas of low detection fraction.}
\label{fig:four}
\end{figure}

We add another layer of complexity in fig. \ref{fig:four}.  We apply our selection function in such a way that the bias of the remaining objects will evolve along the line of sight (the OSF method described in section 
\ref{sec:mocks}).  We continue using the less rare calibrating sample with halo model 
$M_{\rm min}=1\times 10^{12}$ for this illustrative proof of concept.   Although it is not very statistically significant, the eye can pick out a trend in the reconstruction: the reconstructed 
distribution is suppressed in areas of low detection fraction.  Since the bias is tracking the detection 
fraction, the areas of low detection are enhanced less than the areas of high detection, thus they appear
suppressed when we normalize to the highest point.  At this level of accuracy, it is unlikely that this 
systematic will dominate the error in tomographic analyses, however for much larger surveys it could
be a significant concern.

\begin{figure}
\begin{center}
\resizebox{3.7 in}{!}{\includegraphics{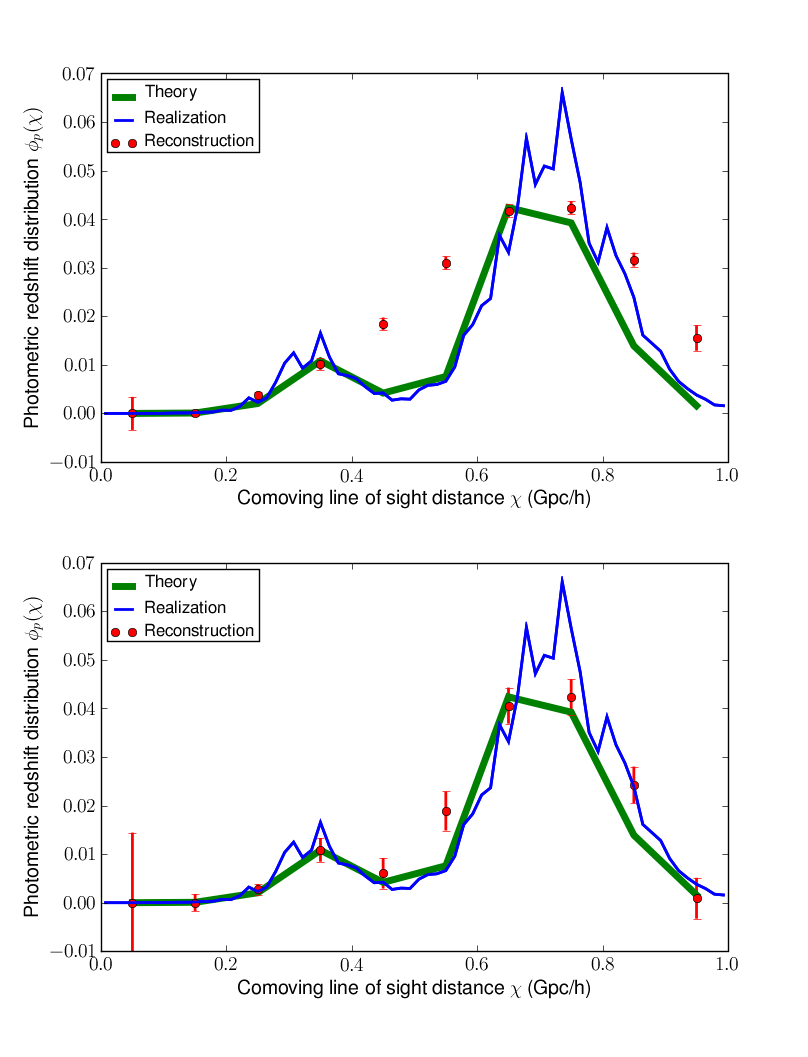}}
\end{center}
\caption{A fair subsample is used to reconstruct $\phi_p(\chi)$ instead of the $M_{\rm min}=1 \times 
10^{12}$ sample.  The bias of the subsample evolves identically to the bias o the photometric sample.  The reconstruction is studied only for the distribution on the right of figs. 
\ref{fig:realist}, \ref{fig:three}, and \ref{fig:four}.  
The top panel shows a subsample of 50\%, which is not sufficient to 
capture the bimodal behavior, and 80\% on the bottom, which does well, and shows no systematic 
offset due to evolving bias. }
\label{fig:five}
\end{figure}

In fig. \ref{fig:five} 
we show how the situation is altered if a fair subsample of the photometric population is 
used instead of a rarer biased tracer population.  This is a special case because the bias of the 
calibrators will evolve with redshift identically to the photometric sample.  
We show two different subsamples in this plot, 50\% and 80\% of the photometric 
catalog, and we test the reconstruction for the distribution on the right in figs. \ref{fig:realist},\ref{fig:three},\ref{fig:four}.  We see that for a survey of this volume, 50\% is too small to capture the bimodal behavior in the reconstruction, but with 80\%, the reconstruction works well.  Indeed the 
systematic offset in figure \ref{fig:four} 
is remedied and the reconstruction follows the realization to much better accuracy.   For larger 
surveys the fraction needed will be smaller than indicated here, because they will be able to measure 
the spectroscopic correlation functions with greater accuracy.  For 
surveys large enough not to be limited by noisy spectroscopic correlation functions, 
it may be necessary to calibrate with a fair subsample of galaxies to avoid
systematic bias in the redshift distribution that comes from evolution in the bias of the photometric sample.

\section{Discussion}\label{sec:disc}
We have outlined the theory behind the cross-correlation method for calibrating the redshift 
distribution of objects with photometric redshifts, and developed a pipeline than can be used
to apply the method to survey data.  We have created mock simulations to test the pipeline.
We have succeeded in reconstructing the redshift distribution of the mock photometric 
galaxies using the angular cross correlation of these galaxies with an overlapping 
spectroscopic sample (whose redshifts are known).   
We have not used any redshift information about the photometric sample.  We have 
demonstrated the validity of the method. 
We have also identified the aspects that are likely to be the limiting factors in relying upon this 
method to provide accurate redshift distribution information.  These limiting factors are  1) that the spectroscopic sample must be binned along the line of sight, causing 
their correlation functions to be noisy and interfering with the reconstruction, and 2) that the bias 
evolution of the photometric sample cannot be disentangled from the redshift distribution that is reconstructed, which may force the necessity for the follow up of a fair subsample of galaxies.  
Improved modeling of theoretical priors could yield large dividends if these factors can be mitigated.   

The analysis has revealed a number of trade-offs that exist in application of this method.   For a 
given set of calibrators, there is a trade-off between the resolution (bin spacing) of the redshift
distribution reconstruction, and the error bars on any individual point.  Thus if a population of 
catastrophic outlier were discovered, for example, it would be interesting to investigate whether 
it is more important to know how many there are, or at exactly 
which redshift they lie.  We also find a trade-off between the number of calibrators and the quality of the
reconstruction.  As the number of available spectra are decreased, the spectroscopic redshift bins 
need to widen to maintain equivalent signal strength.  This in turn affects how finely the reconstructed
redshift distribution can be sampled.  Bimodal behavior may be lost, and with insufficient priors 
(such as the smoothness prior we implemented here), false bimodal behavior may appear in the 
form of anti-correlated adjacent points.  

Widening the bins to compensate for fewer spectra also introduces another difficulty.  Recall that the 
error in the angular cross-correlation function goes as $\delta w_{ps} \sim 1/ \sqrt{n_{\rm pairs}}$.  This means that 
\begin{eqnarray}
\frac{\delta w_{ps}}{w_{ps}}=\frac{1}{\sqrt{n_{\rm pairs}}} \left[ \int_0^\infty \xi_{ps} \phi_p d\chi\right]^{-1}
\end{eqnarray}
For the purposes of this order of magnitude argument, suppose $\phi_{p}$ were constant, then to 
integrate to 1 requires $\phi_{p}(\chi) \propto 1/\Delta\chi_{\rm rb}$.  The subscript rb is used to indicate
the width of the reconstruction bin (not the spectroscopic bins).
When we remove it from the integral we are left with an expression that is $w_p(r_p)$
\begin{eqnarray}
\frac{\delta w_{ps}}{w_{ps}}=\frac{1}{\phi_p \sqrt{n_{\rm pairs}}} \left[ \int_0^\infty \xi_{ps} d\chi\right]^{-1}
\nonumber \\
\sim \frac{\Delta\chi_{\rm rb}}{\phi_p w_p(r_p) \sqrt{n_{\rm pairs}}}
\end{eqnarray}
If the reconstruction bin is widened by a factor of two, the number density of photometric galaxies will have to be increased by a factor of 4 to preserve the same
accuracy in the cross correlation measurement.   Therefore in this method there is also a trade-off between the number of spectra that the survey can afford versus the number of photometric galaxies they can 
afford.  

In this analysis we have not made any use of the photometric redshifts, which although not sufficiently accurate to determine the true redshift distribution of the population, still contain significant information
about it.  Modern techniques such as in \cite{2009arXiv0908.4085G} have made it possible to assign a probability distribution for 
the redshift of each individual galaxy in a photometric survey, rather than a single best estimate and
error bar.   We propose that combining these probabilities for all the galaxies in a given 
redshift bin 
constitutes a reasonably powerful prior, that can take the place of the smoothing we have introduced in this analysis.  This is fortunate because the solution is frequently wrecked by the 
smoothing criterion, although the analysis cannot be performed without it.  Another detail that we leave
for a future study is that the spectroscopic sample need not be comprised of a single rare population, 
it is quite conceivable to target certain regions of redshift space that are known to be problematic more
heavily.  It must also be possible to use less rare tracers in regions with smaller volume.  We leave such
optimization to the future.  

There are a few important factors that we have neglected in this analysis.  As pointed out by \cite{2009arXiv0902.2782B}, 
weak gravitational lensing will induce correlations between the positions of calibrator galaxies in the foreground with photometric galaxies in the background, and vice versa.  This will 
need to be carefully controlled for the method to reliably calibrate redshift distributions.  We also 
have not mentioned the complication of the integral constraint, which as discussed in e.g. 
\cite{2007APh....26..351H} can lead to very significant errors when the volume and the scales in the 
correlation function are of comparable size.  This may be as significant an issue as evolution in the
large scale bias of the photometric population, though it may be mitigated if integral constraint errors
are correlated between photometric and spectroscopic samples.  Fortunately, improved estimators 
exist (in \cite{2007MNRAS.376.1702P} for example) and should certainly be incorporated into the pipeline.  

Having now demonstrated that the method is viable, with further refinement the cross-correlation 
method applied in conjunction with direct follow up surveys may significantly reduce the number of 
spectra that are required to calibrate photometric redshift distributions to the desired accuracy.  There
are a few other advantages that we have not touched upon in detail.  As we have shown, the cross 
correlation method can be used to calibrate the redshift distribution all the way along the line of sight, 
and as such is uniquely suited to detection and calibration of catastrophic redshift errors, even if 
these errors are so rare that they are missed by conventional follow up.  Also, 
the reconstruction should not be adversely affected by redshift deserts, regions where no 
spectroscopic redshifts are available.  We are optimistic that this technique will provide a useful 
complementary approach to conventional calibration techniques, and every effort should be made
to refine the method further. 

\section*{Acknowledgments}
Many thanks to Martin White, for the use of his simulations and for pointing the way out of 
a number of tight corners.  Conversations and e-mails with Rachel Mandelbaum, Jeff Newman, 
Nikhil Padmanabhan, Doug Rudd, William Schulz, David Shih, 
and many other were also incredibly helpful.  A.E. Schulz is 
supported by the Corning Glassworks Foundation Fellowship at the Institute of Advanced 
Study.  

\bibliographystyle{apj}
\bibliography{ms}

\appendix

\section{Error Propagation}\label{app:error}
The treatment in this paper proceeds by assuming that the correlation function 
of the spectroscopic calibrating sample is perfectly determined.  However we show
that since the tracers are rare, the measurements of $\xi_{ss}$ 
can be quite noisy due to binning finely along the line if sight.  Error in the measurement of 
$\xi_{ss}$ may ultimately  dominate the error budget, and it is therefore
important to lay out the procedure for properly incorporating it.  Ignoring the
regularization term, the expression for $\chi^2$ is 
\begin{eqnarray}
\chi^2=({\bm w}-{\bf X}{\bm \phi})^T {\bf C}^{-1}_{\bm \phi}({\bm w}-{\bf X}{\bm \phi})
\end{eqnarray}
${\bf C}_{\bm \phi}$ is the covariance matrix that includes errors in both ${\bm w}$ 
and ${\bf X}$.  The elements of the covariance matrix are
\begin{eqnarray}\label{cov}
{\bf C}_{{\bm \phi},ij}= {\bf C}_{ij}+\sum_k {\bf C'}_{ijk}{\bm \phi}^k +\sum_{k,l}{\bm \phi}^k 
{\bf C''}_{ijkl} {\bm \phi}^l
\end{eqnarray}
which depend explicitly on the solution ${\bm \phi}$.  This renders the problem non-linear
and will require iterative numerical methods to minimize $\chi^2$. 
To cut down on the proliferation of indices, the
expressions in ${\bf C}$ are written out below for the case of a single angular bin $\theta$.  
\begin{eqnarray}\label{eqn:2dc}
{\bf C}_{ij}=\left<\epsilon_w(\chi_i) \epsilon_w(\chi_j)\right> \\ \label{eqn:3dc}
{\bf C'}_{ijk}=\left<\epsilon_w(\chi_i) \epsilon_{\xi}(\chi_j,\chi_k)\right> \\ \label{eqn:4dc}
{\bf C''}_{ijkl}=\left<\epsilon_{\xi}(\chi_i,\chi_j) \epsilon_{\xi}(\chi_k,\chi_l)\right> 
\end{eqnarray}
To generalize to multiple theta bins, let the index $i$  in e.g. $\epsilon_w(\chi_i)$ run over 
each unique combination of $\theta \chi$, rather than just over $\chi$.  
In the expressions for $\epsilon_{\xi}(\chi_i,\chi_j)$ it is worth mentioning that in general $i$ and 
$j$ run over a different number of bins, and $k$ and $l$ run over the number of reconstruction bins
$N$. 

It is often useful in numerical minimization to have an expression for the derivative of $\chi^2$. 
This is
\begin{eqnarray}
\frac{d\chi^2}{d {\bm \phi}}=&2({\bm w}-{\bf X}{\bm \phi})^T {\bf C}^{-1} (-{\bf X}) +({\bm w}-{\bf X}{\bm \phi})^T {\bf C}^{-1} \,\frac{d {\bf C}}{d {\bm \phi}} \, {\bf C}^{-1} 
({\bm w}-{\bf X}{\bm \phi})
\end{eqnarray}
To simplify the expression we have suppressed the indicies, but note that 
the derivative of ${\bf C}$ is a rank 3 object, with two dimensions of the length of ${\bm w}$, 
and one dimension of the length of ${\bm \phi}$ (i.e. $N$).  
Since ${\bf C}$ is a symmetric matrix, the derivative is given by
\begin{eqnarray}
\frac{d {\bf C}}{d {\bm \phi}} = {\bf C'}_{ijk} + 2 \sum_i \phi^i {\bf C''}_{ijkl}
\end{eqnarray}

\end{document}